\documentclass[aip, rsi, amsmath, amssymb, reprint]{revtex4-1}

\usepackage{graphicx}
\usepackage{dcolumn}
\usepackage{bm}
\usepackage{epstopdf}
\usepackage{caption}
\usepackage{subcaption}

\usepackage{dcolumn}

\begin{document}

\title{Increased interference fringe visibility from the post fabrication heat treatment of a perfect crystal silicon neutron interferometer}

\author{B. Heacock}
\email{bjheacoc@ncsu.edu}
\affiliation{Department of Physics, North Carolina State University, Raleigh, NC 27695, USA}
\affiliation{Triangle Universities Nuclear Laboratory, Durham, North Carolina 27708, USA}

\author{M. Arif}
\affiliation{National Institute of Standards and Technology, Gaithersburg, MD 20899, USA}

\author{D. G. Cory}
\affiliation{Institute for Quantum Computing, University of Waterloo, Waterloo, Ontario N2L 3G1, Canada}
\affiliation{Department of Chemistry, University of Waterloo, Waterloo, ON, Canada, N2L 3G1}
\affiliation{Perimeter Institute for Theoretical Physics, Waterloo, ON, Canada, N2L 2Y5}
\affiliation{Canadian Institute for Advanced Research, Toronto, Ontario, Canada, M5G 1Z8}

\author{T. Gnaeupel-Herold}
\affiliation{National Institute of Standards and Technology, Gaithersburg, MD 20899, USA}

\author{R. Haun}
\affiliation{Department of Physics, Tulane University, New Orleans, LA 70118, USA}

\author{M. G. Huber}
\affiliation{National Institute of Standards and Technology, Gaithersburg, MD 20899, USA}

\author{M. E. Jamer}
\affiliation{National Institute of Standards and Technology, Gaithersburg, MD 20899, USA}

\author{J. Nsofini}
\affiliation{Institute for Quantum Computing, University of Waterloo, Waterloo, Ontario N2L 3G1, Canada}
\affiliation{Department of Physics and Astronomy, University of Waterloo, Waterloo, Ontario N2L 3G1, Canada}

\author{D. A. Pushin}
\affiliation{Institute for Quantum Computing, University of Waterloo, Waterloo, Ontario N2L 3G1, Canada}
\affiliation{Department of Physics and Astronomy, University of Waterloo, Waterloo, Ontario N2L 3G1, Canada}

\author{D. Sarenac}
\affiliation{Institute for Quantum Computing, University of Waterloo, Waterloo, Ontario N2L 3G1, Canada}
\affiliation{Department of Physics and Astronomy, University of Waterloo, Waterloo, Ontario N2L 3G1, Canada}

\author{I. Taminiau}
\affiliation{Quantum Valley Investments, Waterloo, Ontario N2L 0A9, Canada}

\author{A. R. Young}
\affiliation{Department of Physics, North Carolina State University, Raleigh, NC 27695, USA}
\affiliation{Triangle Universities Nuclear Laboratory, Durham, North Carolina 27708, USA}

\date{\today}

\begin{abstract}

Construction of silicon neutron interferometers requires a perfect crystal silicon ingot (5~cm to 30~cm long) be machined such that Bragg diffracting ``blades'' protrude from a common base.   Leaving the interferometer blades connected to the same base preserves Bragg plane alignment, but if the interferometer contains crystallographic misalignments of greater than about $ 10 \; \mathrm{nrad}$ between the blades, interference fringe visibility begins to suffer.    Additionally, the parallelism, thickness, and distance between the blades must be machined to micron tolerances. Traditionally, interferometers do not exhibit usable interference fringe visibility until $30 \; \mu \mathrm{m}$ to $60 \; \mu \mathrm{m}$ of machining surface damage is chemically etched away. However, if too much material is removed, the uneven etch rates across the interferometer cause the shape of the crystal blades to be outside of the required tolerances. As a result, the ultimate interference fringe visibility varies widely among neutron interferometers that are created under similar conditions. We find that annealing a previously etched interferometer at $800^\circ \mathrm{C}$ dramatically increased interference fringe visibility from 23~\% to 90~\%.  The Bragg plane misalignments were also measured before and after annealing using neutron rocking curve interference peaks, showing that Bragg plane alignment was improved across the interferometer after annealing.  This suggests that current interferometers with low fringe visibility may be salvageable and that annealing may become an important step in the fabrication process of future neutron interferometers, leading to less need for chemical etching and larger, more exotic neutron interferometers.

\end{abstract}

\maketitle
%
\section{Introduction}
\label{sec:Intro}

\begin{figure}
\centering
\includegraphics[width=0.9 \columnwidth]{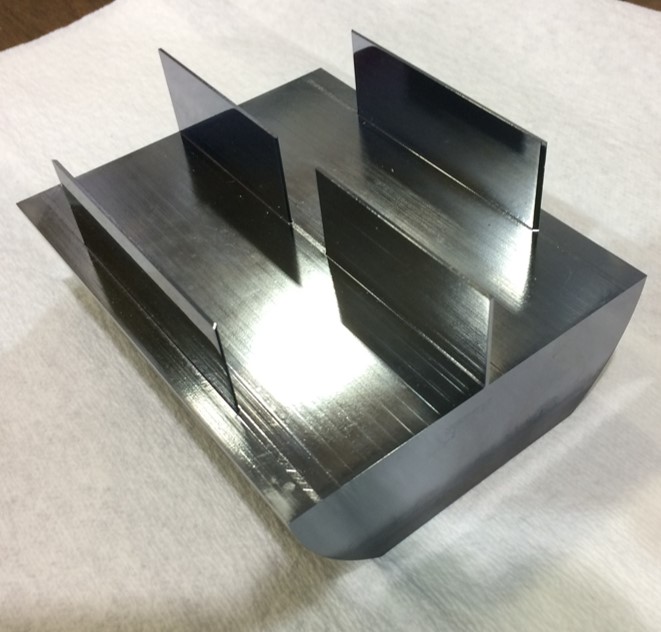}
\caption{The neutron interferometer annealed in this experiment.  The base is roughly 10~cm x 10~cm.  The blades are 3~cm tall.}
\label{fig:ThinBlade}
\end{figure}

\begin{figure}
\centering
	\includegraphics[width=0.88\columnwidth]{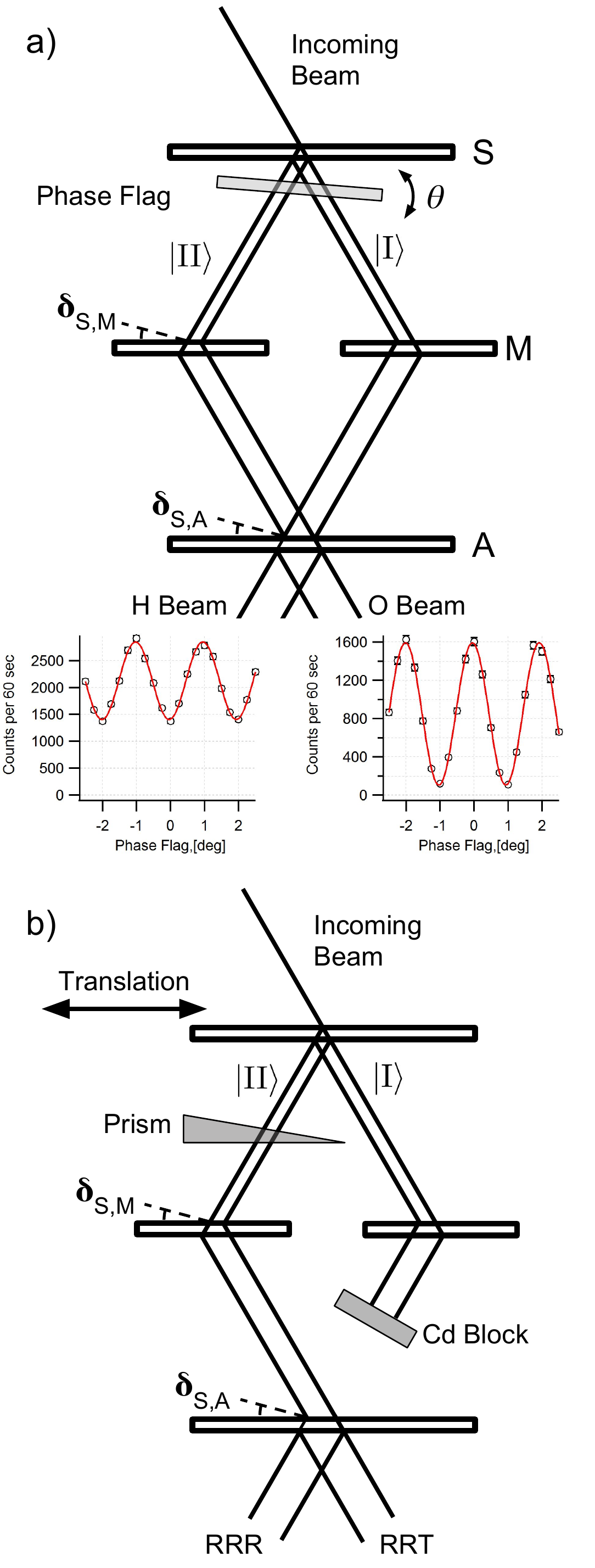}
	\caption{The geometries used in this experiment. Shown in (a), rotating a phase flag, $\theta$, in the interferometer generates a sinusoidal signal.  Shown in (b) is the geometry used to measure the misalignment and thickness variation in the interferometer. The interference peaks upon rotating the prisms about the beam axis are shown in Figs.~\ref{fig:RR} and~\ref{fig:RRR}.  Figures modified from Ref.~\onlinecite{heacock2017neutron}.}
	\label{fig:Geom}

\end{figure}

Perfect silicon neutron interferometers (see Fig.~\ref{fig:ThinBlade}) coherently split and recombine an incoming neutron beam using a series of Bragg diffractions.  The macroscopic separation of the beam paths has led to many historically important experiments over the last forty years, including demonstrations of gravitational quantum interference, the $4 \pi$ periodicity of Dirac spinors, violation of Bell-like inequalities, phase and contrast imaging, neutron holography, and more. \cite{rauch1974test, colella1975observation, rauch1975verification, werner1975observation, hasegawa2003violation, pfeiffer2006neutron, clark2015controlling, sarenac2016holography,lemmel2015neutron,li2016neutron} For a history of the field, see Ref.~\onlinecite{rauch2015neutron}.  The required relative Bragg plane alignment of the splitter, mirror, and analyzer diffracting crystals (labeled S, M, and A, respectively in Fig.~\ref{fig:Geom}) has only ever been achieved by cutting neutron interferometers from a single float zone grown silicon ingot using a rotating diamond saw, leaving splitter, mirror, and analyzer crystal ``blades" protruding from a common base.  The interferometer is then etched in a mixture of hydroflouric, nitric, and sometimes acetic acids. For a good description of interferometer fabrication see Ref.~\onlinecite{zawisky2010large}.  

Neutron interferometers are typically etched iteratively by removing 10's of microns with each etch, then checking contrast by rotating a flat piece of fused silica between the splitter and mirror or mirror and analyzer blades.  The resulting sinusoidal neutron interference signal can theoretically have 100~\% contrast in the O-Beam (Fig.~\ref{fig:Geom}a), where the contrast is given by the amplitude over the mean of the fitted oscillation.  Etching is believed to relieve strain in the crystal caused by machining damage. \cite{zawisky2010large}  However, as the total etching depth increases, the parallelism and uniform thicknesses of the crystal blades degrade due to uneven etch rates.  If too much material is etched away, the contrast begins to drop.  Additionally, it is well established that neutron interferometer contrast varies depending on where the incident beam strikes the splitter blade and for different wavelengths.  We recently measured a variation in Bragg plane alignment across a 1~cm span of the splitter blade of another interferometer of up to 40~nrad,~\cite{heacock2017neutron} a phenomenon suggested earlier by Ref.~\onlinecite{arthur1985dynamical}.  This finding is confirmed here, as well as a variation in the blade thicknesses across the particular interferometer used in this experiment.  

Interferometers constructed under similar conditions can show wildly different contrasts.  Previously, machining accuracy was thought to be a major source of this variation.  However, modern machining processes eliminate this as a possibility.  This work demonstrates that the lack of reproducibility in interferometer construction is likely due to thickness variations of the interferometer blades from uneven etch rates as well as fluctuating or large (greater than 10~nrad) Bragg plane misalignments between the blades.  We show that Bragg plane misalignment can be reduced by annealing the interferometer after fabrication.



\section{Bragg Diffraction and the Neutron Coherence Length}

Because neutron beams have a spread in momentum space that is much broader than the angular acceptance for Bragg scattering from a perfect silicon crystal, called the Darwin width $\Theta_\mathrm{D}$, the coherence length of a Bragg-diffracted neutron wavepacket along the diffraction direction is given by the pendell\"{o}sung length, which is about 50 microns for typical neutron wavelengths and first order Bragg diffraction.  If machining inaccuracies or uneven etch rates cause the two beam paths in a neutron interferometer (Paths $|\mathrm{I} \rangle$ and $ | \mathrm{II} \rangle$ in Fig.~\ref{fig:Geom}a) to be displaced relative to each other outside of the pendell\"{o}sung length, then the interferometer contrast suffers.  This sets the machining tolerances for the interferometer blades, indicating that the parallelism, thickness, and spacing of the crystal blades should be uniform to a level much smaller than the pendell\"{o}sung length. \cite{rauch2015neutron}

Bragg plane misalignments between diffracting blades of the interferometer can also cause contrast to suffer due to details of neutron dynamical Bragg diffraction by a perfect crystal.  (For a complete description of dynamical diffraction see Refs.~\onlinecite{sears1989neutron, rauch2015neutron,lemmel2013influence}.  Dynamical diffraction can also be approached from a quantum information perspective. \cite{nsofini2016quantum,nsofini2017noise})  For subsequent diffracting crystals of the same thickness, interference peaks appear when all the phases of the momentum states making up the neutron wavepacket interfere constructively in the twice, or more, reflected beam.  These peaks can be measured by rotating a refracting prism about the beam axis between blades of an interferometer.  The prism slightly deflects the beam, and rotating the prism moves the refraction plane of the prism in and out of the diffraction plane of the crystal.  \cite{bonse1977oscillatory, pet1984multiple, arthur1985dynamical} These interference peaks have an angular scale in the Bragg plane misalignment of $\delta = (H D)^{-1} \sim 100 \; \mathrm{nrad}$, where $H$ is the reciprocal lattice vector, and $D$ is the crystal thickness.  The same dephasing effect occurs in Mach-Zhender neutron interferometers, thus setting the misalignment tolerances of the Bragg planes in each blade of a neutron interferometer.  For a generalized description of how interferometer contrast is affected by Bragg plane misalignments, see Ref.~\onlinecite{heacock2017neutron}.  Additionally, a computational program outlined in Refs.~\onlinecite{lemmel2007dynamical, potocar2015neutron} is able to accommodate thickness differences and different incoming momentum space beam profiles in a neutron interferometer.


To characterize the impact of annealing on Bragg plane alignment of the interferometer blades, we measured the interference structure of multiple Bragg diffractions in one arm of the interferometer as a function of rotation of a fused silica prism. \cite{bonse1977oscillatory, pet1984multiple, arthur1985dynamical, heacock2017neutron} In this way, we can directly see the relative Bragg plane alignment of the interferometer blades change by about $100 \; \mathrm{nrad}$ after annealing as a shift in interference peak position.  The structure of each interference peak is perturbed by unequal crystal thicknesses on the pendell\"{o}sung length scale, and we find that the overall structure of each peak does not change with annealing.  We are thus able to differentiate between the detrimental effects of unequal blade thicknesses and Bragg plane misalignments in the interferometer and show that the post fabrication annealing of the interferometer can improve the latter.

\section{Experiment}

This work was performed at the National Institute of Standards and Technology (NIST) Center for Neutron Research (NCNR).  There are two dedicated interferometry beamlines at the NCNR.  One beamline uses a 2.7~\AA \; neutron wavelength and a sophisticated vibration isolation system, the details of which can be found in Refs.~\onlinecite{nico2005fundamental, rauch2015neutron}.  The second beamline has 2.2~\AA \; and 4.4~\AA \; neutron wavelengths available.  A description of the second beamline can be found in Refs.~\onlinecite{shahi2016new,pushin2015neutron}.  Both beamlines and all three wavelengths were used in this experiment.

\begin{figure}
\centering
\includegraphics[width=0.9 \columnwidth]{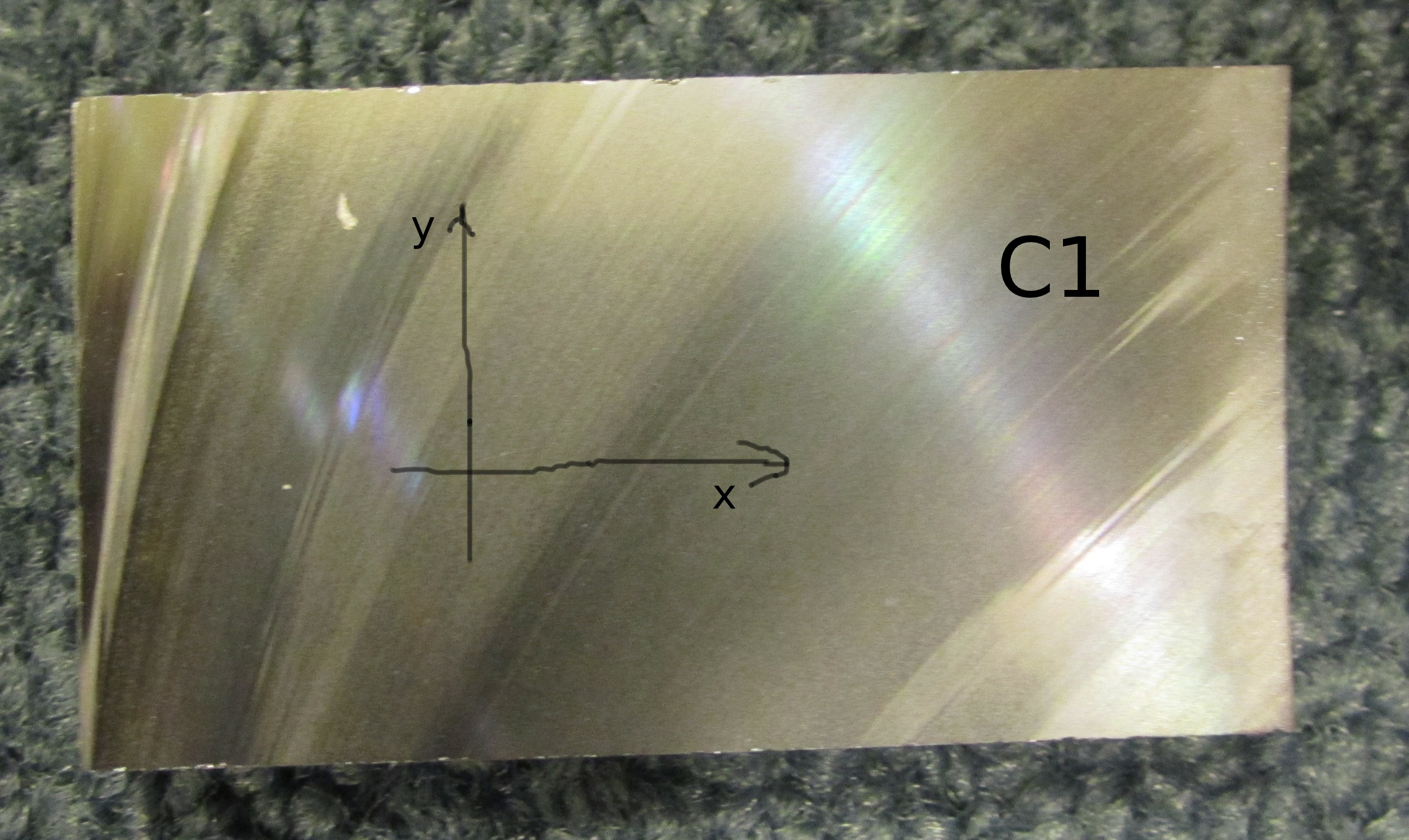}
\caption{The sample crystal \#1.  The coordinate scale denotes the surface stress analysis orientation. }
\label{fig:sample}
\end{figure}

To begin, smaller float-zone (Fz) grown, silicon-crystal samples (see Fig.~\ref{fig:sample}) were annealed at a variety of temperatures before attempting to anneal the much larger interferometer crystal.   The samples were cut using a rotating diamond saw with fine grit size.  None of the samples where etched but several of the samples where further refined using diamond turning machining where micron size cuts could be made. These samples were tested with both x-rays and neutrons.   Unlike silicon crystals grown by the Czochralski (Cz) method, whose oxygen content causes strain in the crystal structure to increase with annealing, crystal planes become more highly ordered with annealing for Fz crystals. \cite{newman1982defects}  X-ray stress analysis of the crystal surfaces was performed on several of the samples using the technique described in Ref.~\onlinecite{reimers1992investigations}.  These results are summarized in Table \ref{tab_sample_anneal} and one can see the reduction of surface stresses in the crystals after annealing. 

The samples where annealed in a tabletop tube furnace.  The annealing process consisted of placing the sample inside a 25.2 mm diameter quartz tube on end of which was connected to a vacuum pump.  The quartz was inserted in the tube furnace and the temperature was ramped at a rate of 1 $^{\circ}$C/min up to a set constant temperature.  The sample was held at the set temperature for several hours then ramped down at the same 1 $^{\circ}$C/min rate. The interferometer itself was annealed in a larger tube furnace with an inner diameter of 208 mm under constant argon flow at $800^\circ \mathrm{C}$ for ten hours, not including ramping time.  The ramping rate was 5~$^\circ$C/min.  Lapped Fz silicon crystal slabs were annealed in the furnace under the same conditions before attempting to anneal the interferometer.  The full width half maximum of an untreated and an annealed sample is shown in Table~\ref{table:XrayData} with the x-ray results from the smaller furnace.

\begin{table*}{\caption{The results of the x-ray surface stress analysis  $\sigma_{xx}$ and  $\sigma_{yy}$ on several test samples.  A  high quality  perfect silicon crystal, which had been etched by 50 microns, was used for comparison. If an annealing temperature is not provided, the measurement was made before the crystal was annealed.  The uncertainties are indicated in the brackets.  The full width half maximum (FWHM) of the reflected x-ray beam intensity is also compared.  Narrower FWHM are indicative of better quality crystals. The stress measurements were performed with a x-ray diffractometer.  The FWHM measurements were performed using a separate diffractometer with a silicon monochromator, except for those marked $^*$, which were performed on a third diffractometer without a crystal monochromator. \label{tab_sample_anneal} }
        \label{table:XrayData}}
		\centering 
		\begin{tabular}[c]{lc@{\hspace{0.25in}}c@{\hspace{0.25in}}c@{\hspace{0.25in}}c@{\hspace{0.25in}}c@{\hspace{0.25in}}r}\hline\hline 
     	Sample	       & Description     & $\sigma_{xx}$ & $\sigma_{yy}$ & Anneal $T$ & FWHM  \\  
              	       &                 & MPa         &  MPa & $^{\circ}$C & arcsec \\ \hline 
        Perfect Crystal& Etched       & -	    & - & -  & 6.96(5)\\
		Crystal \#1    & Diamond saw  & 2(3)    & -121(3) &  -&  \\
        Crystal \#1    & Diamond saw  & -10(3)	&   12(5) & 900& 34.6(5)\\
        Crystal \#2    & Diamond saw  & -       &   - & - & 41.7(8)\\
		Crystal \#4    & Diamond turned & -13(3)&   43(5) & - & 21.4(6)\\
		Crystal \#3    & Diamond saw  & -1(4)   &   17(9) & 700& \\
		Crystal \#3    & Diamond saw  &  6(3)   &    6(5) & 900& 26.7(5)\\	
        Crystal \#5		& Lapped		& - 	&  -	&  - & $150(1)^*$  \\ 
        Crystal \#6		& Lapped		& -		& - 	&800 & $109(1)^*$ \\ \hline
		\end{tabular}

\end{table*}

\begin{figure}
\centering
\begin{subfigure}[b]{0.95\columnwidth}
\includegraphics[width=\textwidth]{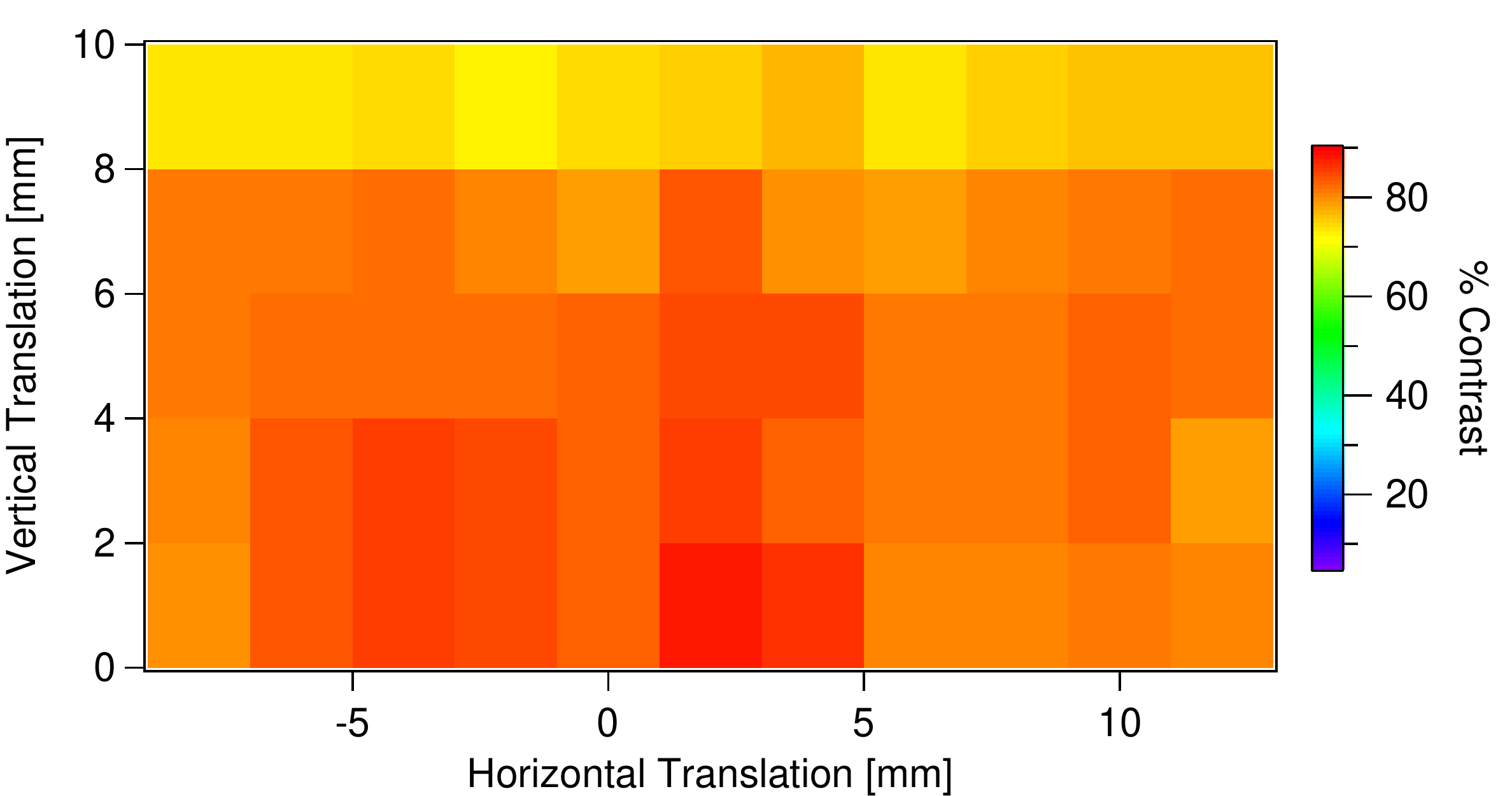}
\caption{}
\label{fig:Map:Ann}
\vspace{3ex}
\end{subfigure}
\begin{subfigure}[b]{0.95\columnwidth}
\includegraphics[width=\textwidth]{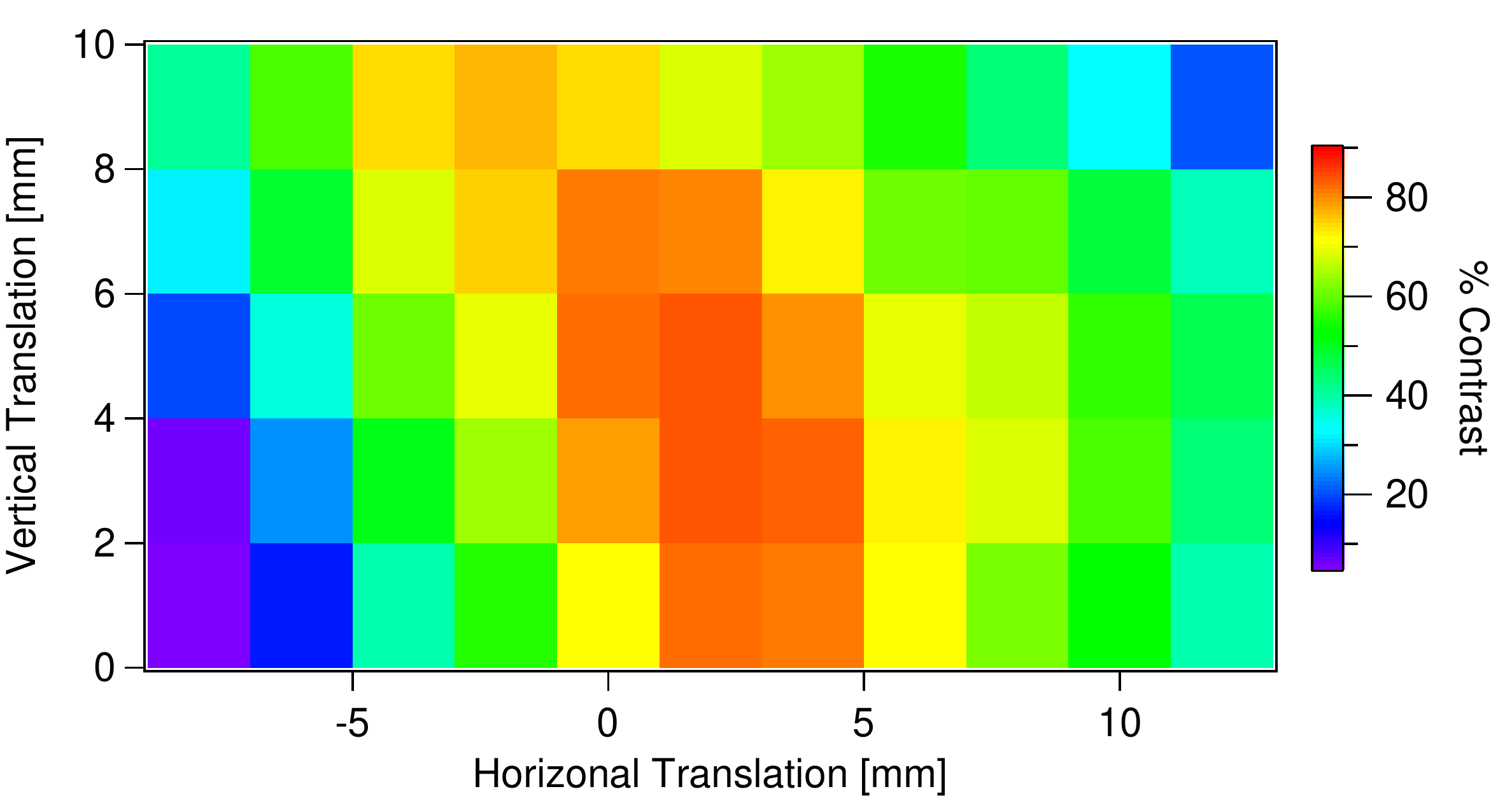}
\caption{}
\label{fig:Map:LLL}
\end{subfigure}
\caption{Contrast Maps at 2.7\AA \; for (a) the annealed interferometer and for (b) the previously highest-contrast NIST interferometer.  The incoming beam passes through a 2~mm x 8~mm slit in both cases.  The peak contrast of the annealed interferometer is only slightly higher, but it shows high contrast over a wider spatial range.}
\label{fig:Map}
\end{figure}

The interferometer that was annealed uses the (111) Bragg reflection. In the past, the interferometer had exhibited maximum contrast of 23~\% at 2.7~\AA. \cite{wood2014quantum}  Immediately before annealing, it was tested for contrast at 2.2~\AA \; and 4.4~\AA, though no visible contrast was found.  When searching for contrast, an interferometer is translated vertically and horizontally along the splitter blade thus making a ``contrast map"  (see Fig.~\ref{fig:Map}).  

Before annealing, the interference structure of misaligning the mirror and analyzer blades, relative to the splitter, was studied by placing a fused silica prism between the splitter and mirror blades of the interferometer.  The prism was placed so that its $6^\circ$ pitch was oriented at a right angle to the diffraction plane to within a few degrees.  By then rotating the prism about the beam axis, deflection of the beam from the prism enters the diffraction plane, which causes the same effect as rotating the analyzer and mirror blades relative to the splitter blade at the nanoradian level.  The structure associated with this rocking curve has been studied in the past. \cite{bonse1977oscillatory, pet1984multiple, arthur1985dynamical}  

The angular deflection in the diffraction plane caused by the prism is

\begin{equation}
\delta = \frac{\lambda^2 }{ 2 \pi} \tan{\alpha} \sin{\phi} \sum_i N_i b_i ,
\label{eqn:deflection}
\end{equation}

\noindent
where $\lambda$ is the neutron wavelength; $\alpha$ is the pitch of the prism; $\phi$ is the tilt of the prism about the beam axis; and the sum is over the number densities $N_i$ and scattering lengths $b_i$ of each species.

As the prism is rotated between the splitter and one of the mirror crystals with the other beam blocked, the reflected-reflected-transmitted (RRT) and reflected-reflected-reflected (RRR) beams are counted in $^3 \mathrm{He}$ detectors (Fig.~\ref{fig:Geom}b).  The position of the RRR peak is an average of the misalignment between the splitter and mirror blades and the splitter and analyzer blades.  By adding the RRT and RRR beams together, we form the reflected-reflected (RR) beam, whose position is given by the Bragg plane misalignment between the splitter and mirror blades.  

To describe the RR and RRR peaks, first we define two special functions:

\begin{eqnarray}
\mathcal{I}(\alpha, \beta) = && \int^1_{-1} d \Gamma \sqrt{1-\Gamma^2} \cos \left ( \alpha \Gamma \right ) \cos \left ( \beta \over \sqrt{1-\Gamma^2}  \right )
\label{eqn:I} \\
\mathcal{J} (\alpha, \beta) = && \int^1_{-1} d \Gamma  \left (  1- \Gamma^2 \right)^{3 \over 2} \cos \left ( \alpha \Gamma  \right) \nonumber \\
&& \times \cos \left ( {\beta \over \sqrt{1-\Gamma^2}} \right ).  \label{eqn:J}
\end{eqnarray}

The RR peak before and after annealing was fit to:

\begin{equation}
I_\mathrm{RR} = A \left \{  \pi + \mathcal{I} \left [ B \left ( \delta(\phi) + \delta_{S,M} \right ), \; \Delta_{M,S} \right ] \right \} ,
\label{eqn:RRfit}
\end{equation}

\noindent
where $A$, $B$, $ \Delta_{M,S}$, and $\delta_{S,M}$ are fit parameters; $\delta (\phi)$ is given by Eqn.~\ref{eqn:deflection} as a function of the the prism rotation $\phi$; and $\delta_{i,j}$ is interpreted as the angular misalignment of the splitter, mirror, or analyzer blades (subscripts $S$, $M$, or $A$, respectively).  Here $\Delta_{M,S}$ is given by the thickness difference of two blades scaled by the pendell\"{o}sung length

\begin{equation}
\Delta_{i,j} = 2 \pi {D_i - D_j \over \Delta_H} ,
\label{eqn:Delta}
\end{equation}

\noindent
where $\Delta_H$ is the pendell\"{o}sung length, and the subscripts again refer to the crystal blades. The RRR peak was fit to

\begin{widetext}
\begin{eqnarray}
I_\mathrm{RRR} = &&A'\Bigg \{ {9 \over 16} \pi + \mathcal{J}  \left [  B' \left (  \delta(\phi) + \delta_{S,M} \right ), \; \Delta_{M,S} \right ]
+ \mathcal{J} \left [  B' ( \delta_{S,M} - \delta_{S,A} ), \; \Delta_{A,S} - \Delta_{M,S} \right ] \nonumber \\
&& + \mathcal{J} \left[ B' (\delta(\phi) + \delta_{S,A}) , \; \Delta_{A,S} \right ] \Bigg \} ,  \label{eqn:RRRfit}
\end{eqnarray}
\end{widetext}

\noindent
where $A'$, $B'$, $\Delta_{A,S}$, and $\delta_{S,A}$ are fit parameters, and $\Delta_{M,S}$ and $\delta_{S,M}$ are taken from the best fits of the RR data (Eqn.~\ref{eqn:RRfit}).  Both $\delta_{S,M}$ and $\delta_{S,A}$ carry the same constant offset from the unknown offset in the prism rotation.  All the fits had $21-4$ degrees of freedom; the reduced $\chi^2$ were between 0.64 and 1.1.

\begin{figure}

\centering
	\begin{subfigure}[b]{0.9\columnwidth}
		\includegraphics[width=\textwidth]{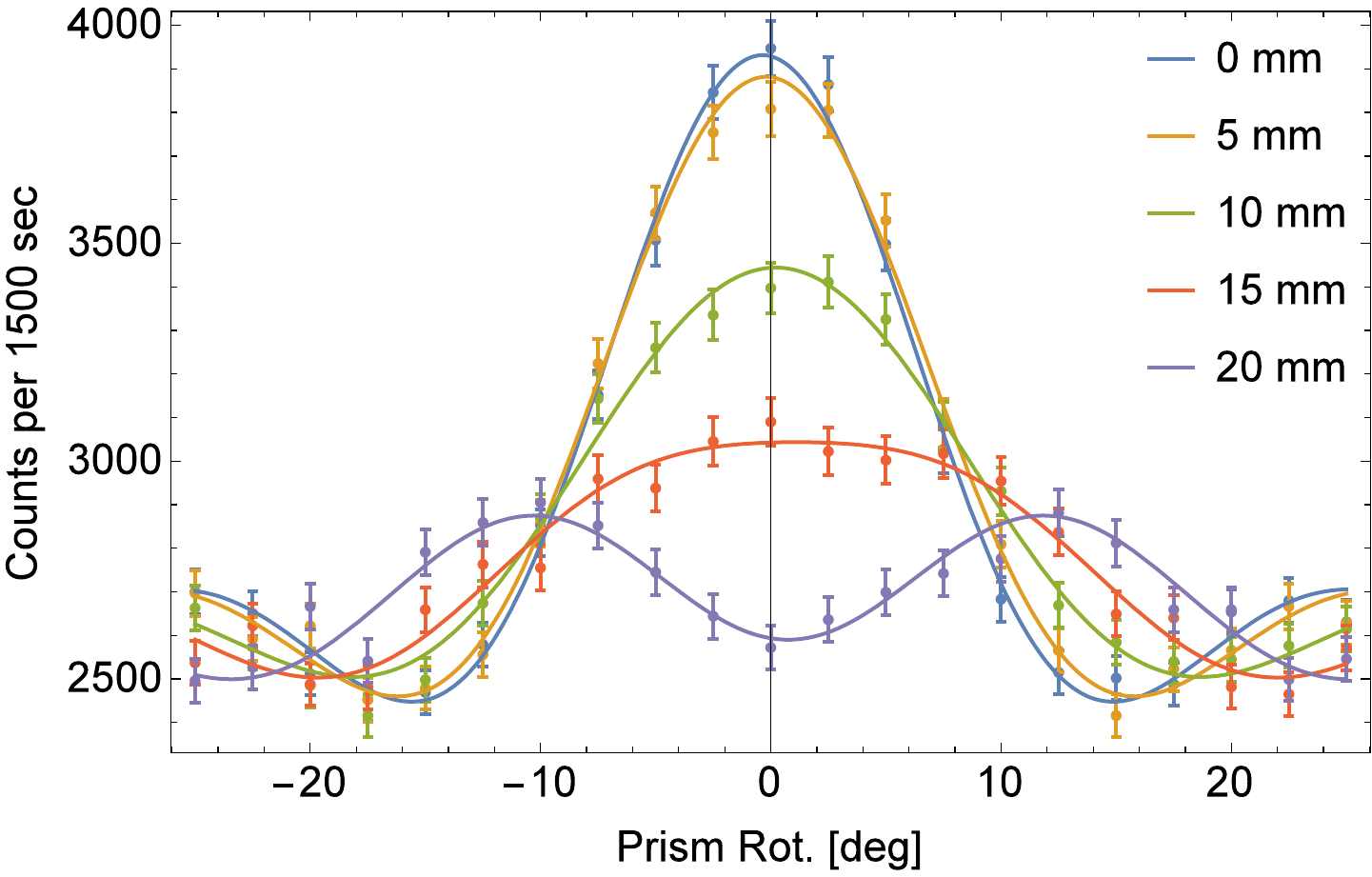}
		\caption{Before Annealing}
		\label{fig:RR:Pre}
	\end{subfigure}
	\\ \vspace{5ex}
	\begin{subfigure}[b]{0.9\columnwidth}
		\includegraphics[width=\textwidth]{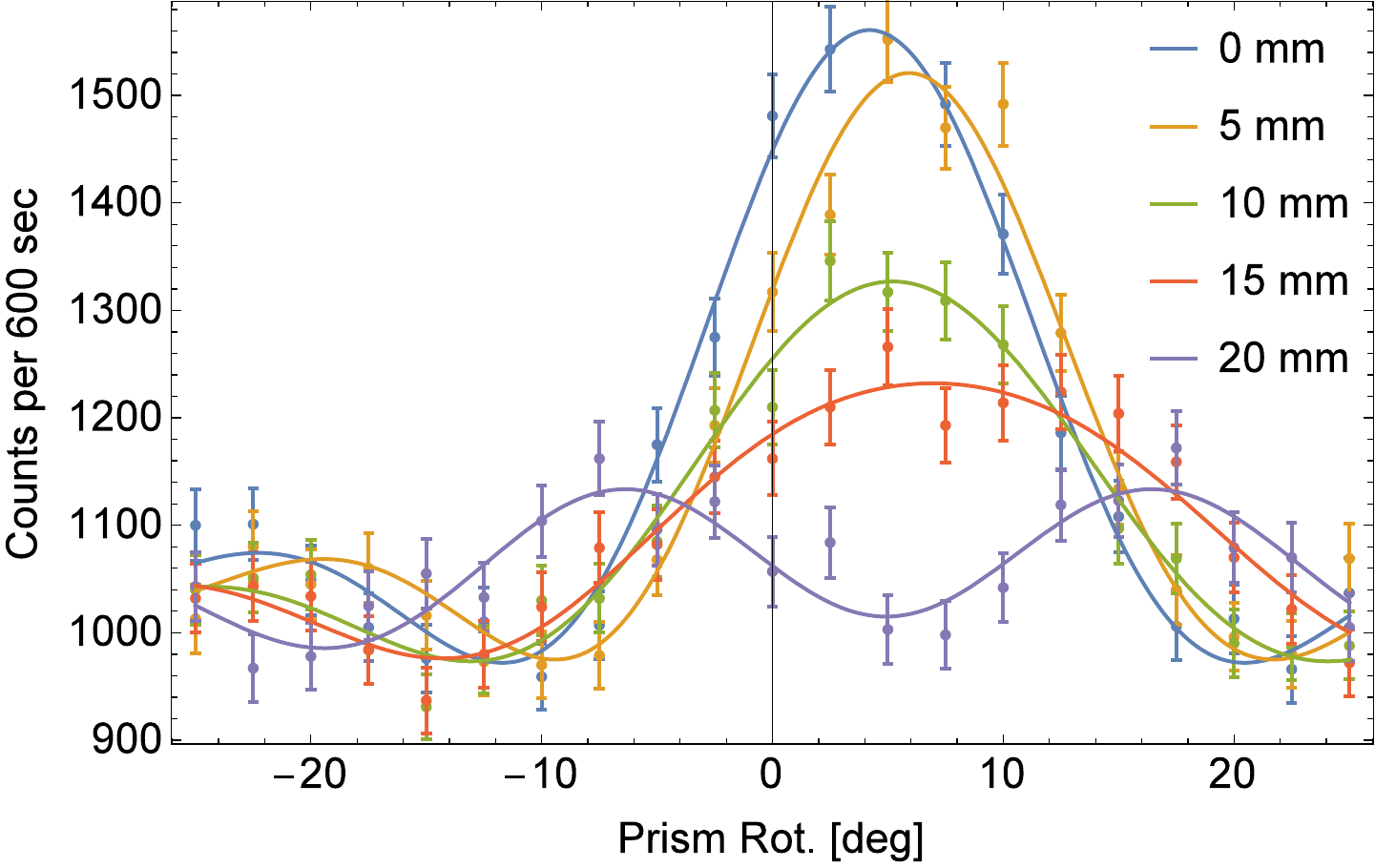}
		\caption{After Annealing}
		\label{fig:RR:Annealed}
	\end{subfigure}
 	\caption{The RR interference peaks before and after annealing with best fits. The visible shift in the curves for (a) before annealing and (b) after annealing is a measurement of the change in Bragg plane alignment from annealing.  While the absolute alignment of the prism rotation is unknown, it was not changed after annealing.  Each curve is for a different translation of the interferometer relative to the incoming beam.}
	\label{fig:RR}
\end{figure}

\begin{figure}

\centering
	\begin{subfigure}[b]{0.9\columnwidth}
		\includegraphics[width=\textwidth]{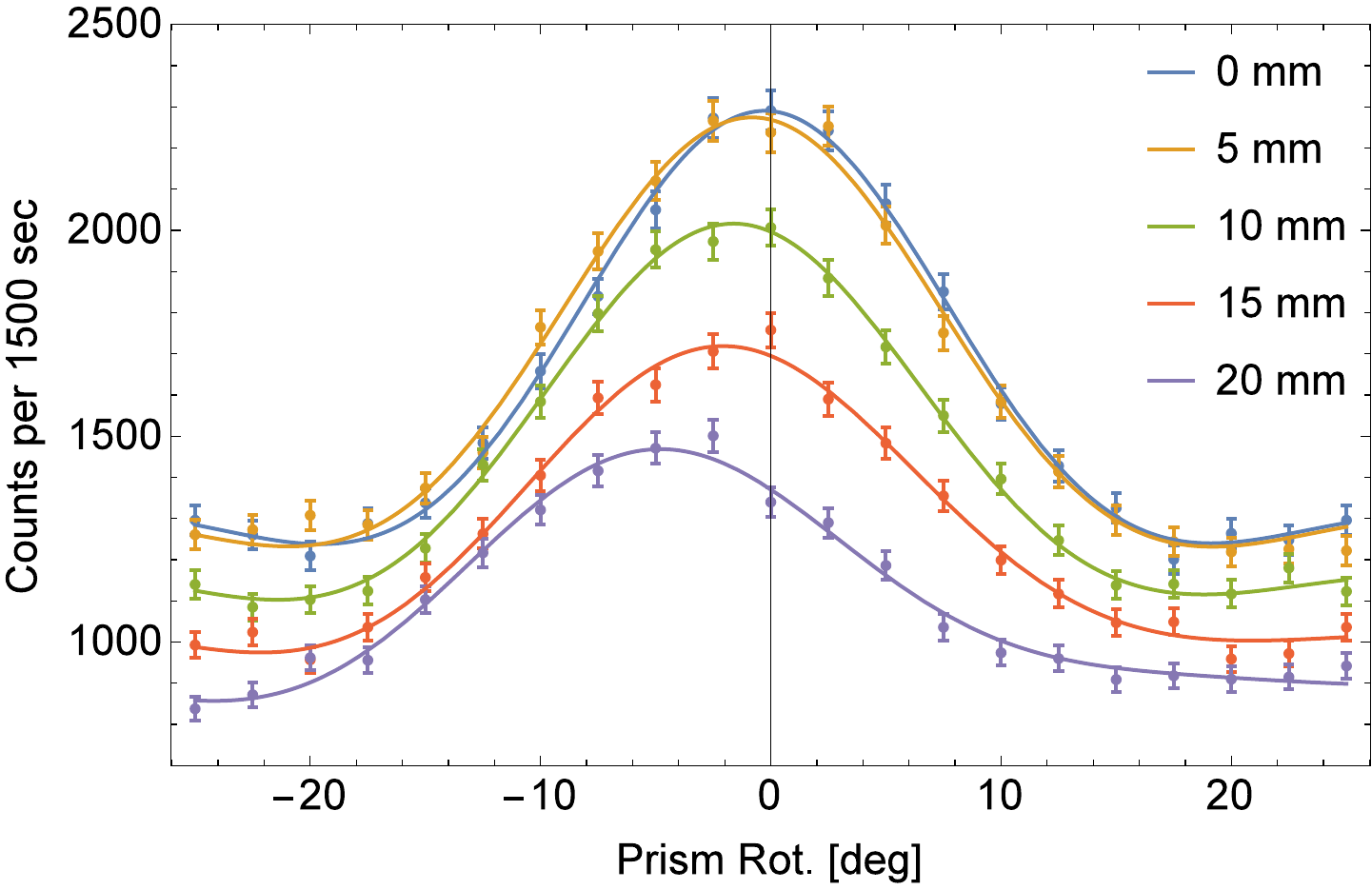}
		\caption{Before Annealing}
		\label{fig:RRR:Pre}
	\end{subfigure}
	\\ \vspace{5ex}
	\begin{subfigure}[b]{0.9\columnwidth}
		\includegraphics[width=\textwidth]{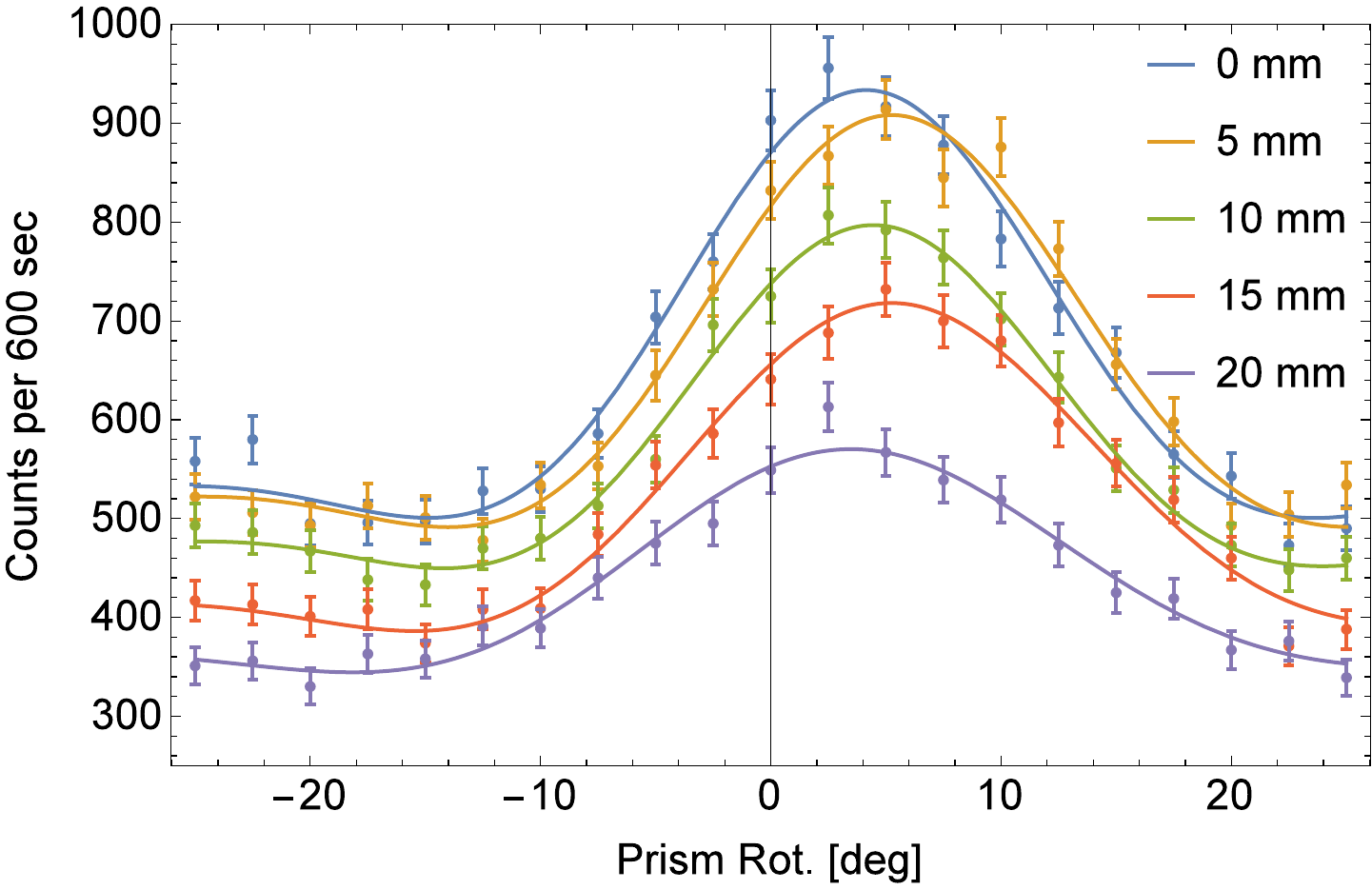}
		\caption{After Annealing}
		\label{fig:RRR:Annealed}
	\end{subfigure}
	\caption{The RRR interference peaks before and after annealing with best fits. The visible shift in the curves for (a) before annealing and (b) after annealing is a measurement of the change in Bragg plane alignment from annealing.  While the absolute alignment of the prism rotation is unknown, it was not changed after annealing.  Each curve is for a different translation of the interferometer relative to the incoming beam.}
	\label{fig:RRR}
\end{figure}

Figs.~\ref{fig:RR} and~\ref{fig:RRR} show the fitted functions for a few incident beam positions on the interferometer at 4.4~\AA.  From these fits the misalignment of each of the crystal blades relative to the splitter plus any offset from the prism rotation is found. Additionally, the structure of the peak lends information on the difference in thickness between the interferometer blades through the $\Delta_{i,j}$ fit parameters.  This effect can be clearly seen in the widening and double peak structure with in Fig.~\ref{fig:RR} as the interferometer is translated.

\begin{figure}
\centering
	\begin{subfigure}[b]{0.9\columnwidth}
		\includegraphics[width=\textwidth]{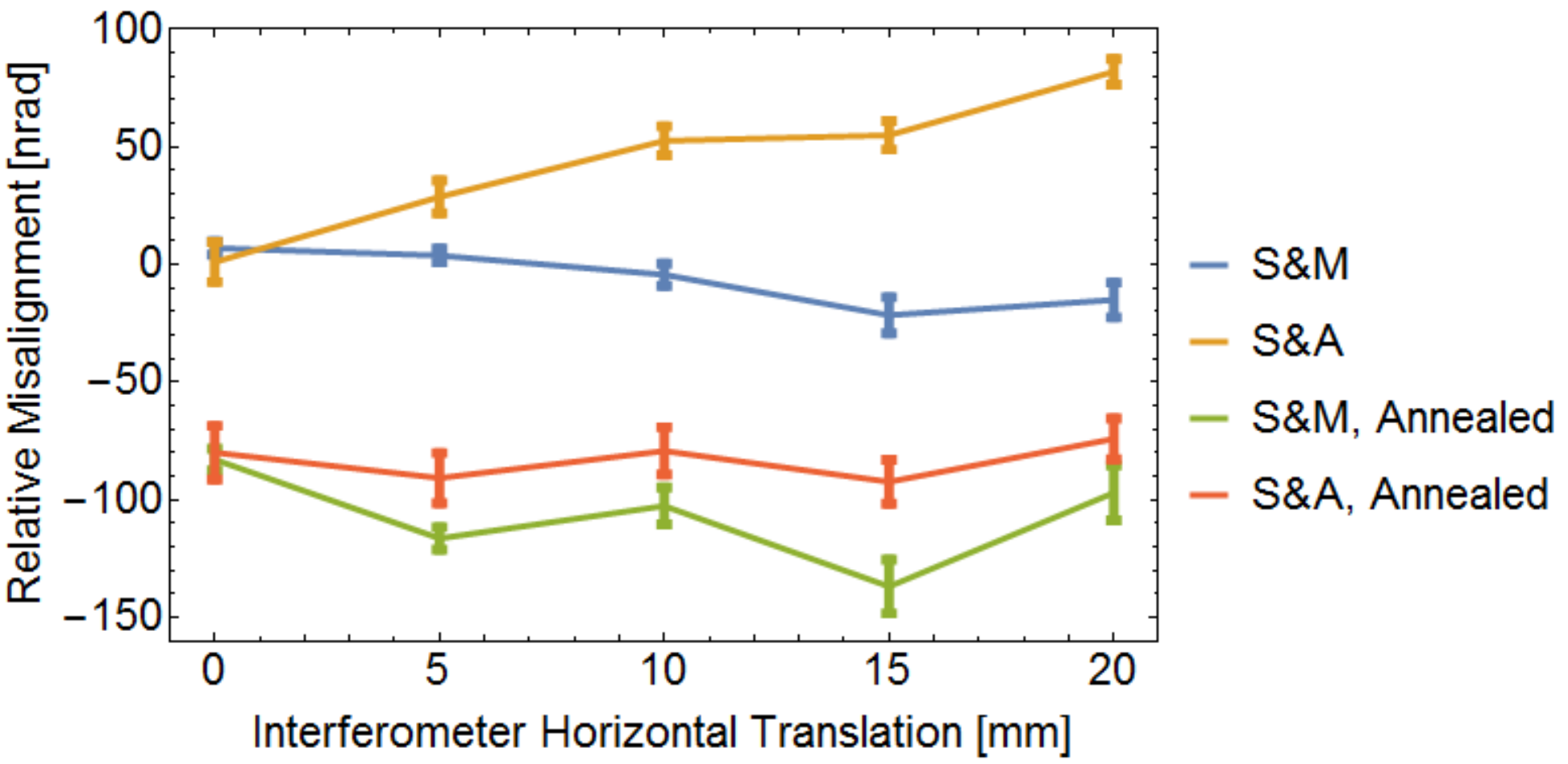}
		\caption{}
		\label{fig:MisThick:Mis}
	\end{subfigure}
	\\ \vspace{5ex}
	\begin{subfigure}[b]{0.9\columnwidth}
		\includegraphics[width=\textwidth]{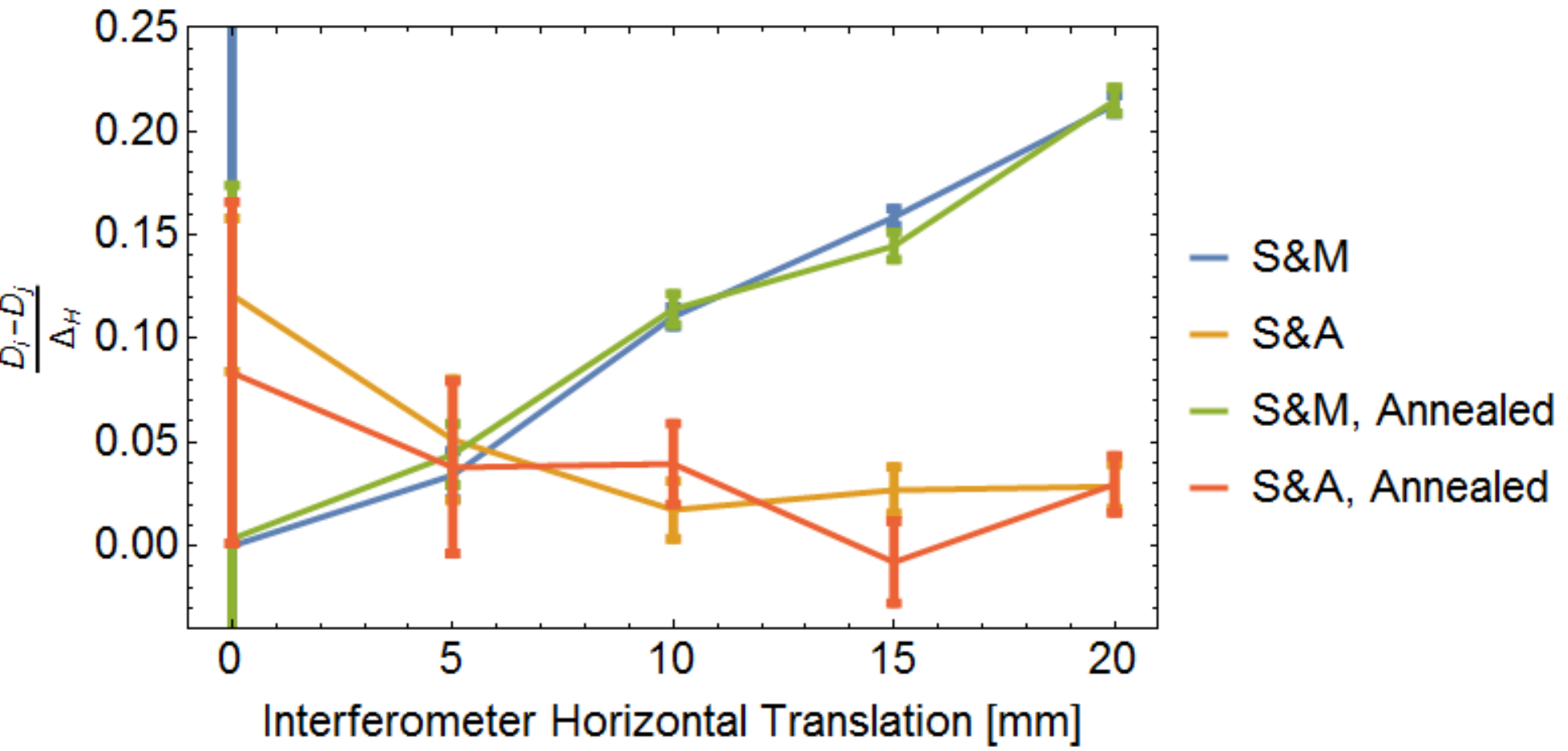}
		\caption{}
		\label{fig:MisThick:Thick}
	\end{subfigure}
	\caption{Bragg plane misalignments and thickness differences before and after annealing.  Shown in (a), Bragg-plane misalignments as a function of interferometer translation.  Note that the y-axis is relative, with a constant offset from the prism rotation.  Shown in (b) are thickness differences as a function of interferometer translation.  $S$, $M$, and $A$ refer to the splitter, mirror, and analyzer blades. }
	\label{fig:MisThick}
\end{figure}

The Bragg plane misalignments before and after annealing are shown in Fig.~\ref{fig:MisThick:Mis}.  Gradients in the Bragg plane misalignments drop from about 10 nrad/mm to less than 5~nrad/mm after annealing.  The fused silica prism was left in place while the interferometer was annealed.  This allowed us to measure a shift in the Bragg plane alignment of the splitter blade and the mirror and analyzer blades of about 100~nrad from annealing, despite the absolute alignment of the prisms being unknown.

The fitted $\Delta_{i,j}$ parameters can be seen as the worsening visible distortion in the RR peak, and less so for the RRR peak, as the interferometer is translated (Figs.~\ref{fig:RR} and~\ref{fig:RRR}).  This indicates that the thickness difference of the mirror and splitter blades changes by about 20~\% of the pendell\"{o}sung length over 20~mm; this corresponds to about a $7~\mu \mathrm{m}$ thickness difference, given the $34~\mu \mathrm{m}$ pendell\"{o}sung length for the (111) reflection at 4.4~\AA.  The same is not true of the RRR peak, indicating that the mirror blade likely has a varying thickness.  Alternatively, the mirror blade could be flat with the splitter and analyzer blades having a similar, distorted shape.  

The technique of using the deflecting prism to measure Bragg plane misalignments only applies to one of the mirror blades, leaving the other unmeasured.  We therefore cannot predict the contrast solely from the fits of the rocking curve interference peaks. The interferometer contrast is only a function of the difference in thickness between the splitter and analyzer blades and the two mirror blades separately.  However, the mean count rate of an interferogram is at its highest when all four diffracting crystals have the same thickness.

Before annealing, there was no contrast visible at 2.2~\AA \; or 4.4~\AA.  After annealing, the interferometer was tested at 4.4~\AA \; and 2.7~\AA. There was up to 20~\% contrast observed at 4.4~\AA \; after annealing.  The contrast at 2.7~\AA \; was excellent, improving from 23~\% \cite{wood2014quantum} to 90~\%.  A contrast map is shown in Fig.~\ref{fig:Map}.  Also shown in Fig.~\ref{fig:Map} is a contrast map of the previously highest-contrast interferometer at NIST.  While the peak contrasts are similar, the range over which the contrast is high is larger for the the annealed interferometer.  This implies that the annealed interferometer may be especially useful for phase imaging, where the incoming beam is much larger (typically $\sim 1 \; \mathrm{cm} $ in diameter).

The better contrast at 2.7~\AA, \; compared to 4.4~\AA, \; for the annealed interferometer is likely due in part to the better vibrational and environmental isolation provided by the different facility.  The more severe Bragg angle at 4.4~\AA \; (44.5$^\circ$ versus 25.6$^\circ$), also creates a larger path separation, rendering the interferometer more sensitive to vibrations.  However, the dependence of contrast on wavelength may also be due in part to thickness variations across the interferometer blades. It is conceivable that the demonstrated thickness variation in the mirror blade that we measured at 4.4~\AA \; (Fig.~\ref{fig:MisThick:Thick}) is less pronounced for the 2.7~\AA \; beam geometry.

\section{Conclusions}

We have shown that annealing a neutron interferometer can refine Bragg plane misalignments enough to drastically improve contrast.  While the Bragg plane alignment can be improved by annealing, the only way to fix thickness variations in the interferometer blades would be to remachine and etch the interferometer again.  The variation in blade thickness is believed to be a principle cause of the lower contrast at 4.4~\AA, \; when compared to 2.7~\AA \; for the interferometer annealed in this work.  It is possible that with the addition of annealing treatments to the neutron interferometer post-machining fabrication process, less etching will be required.  If this is the case, then the annealing step may also prevent thickness variation in the crystal blades caused by deep etching depths, resulting in higher quality interferometers with blades that are more uniform and parallel and that have better Bragg plane alignment.

\acknowledgements

The authors would like to thank Juscelino Leao for helping us with the annealing furnace and John Barker for discussing the annealing of silicon with us.  This work was supported in part by the US Department of Energy under Grant No. DE-FG02-97ER41042, National Science Foundation Grant No. PHY-1307426 and Grant No. PHY-1205342, Canadian Excellence Research Chairs (CERC) (215284), the Canada  First  Research  Excellence  Fund  (CFREF), Natural Sciences and Engineering Research Council of Canada (NSERC) Discovery (RGPIN-418579), Collaborative Research and Training Experience (CREATE) (414061), and the NIST Quantum Information Program.

\end{document}